\documentclass[showpacs,aps,graphicx,twocolumn]{revtex4}
\usepackage{amsmath}
\usepackage{amscd}
\usepackage{graphicx}
\begin{document}

\title{Multipartite entanglement purification with quantum nondemolition detectors\footnote{Published in
Eur. Phys. J. D \textbf{55}, 235-242 (2009)}}

\author{Yu-Bo Sheng$^{1,2,3}$, Fu-Guo Deng$^{4}$\footnote{
e-mail: fgdeng@bnu.edu.cn}, Bao-Kui Zhao$^{1,2,3}$, Tie-Jun
Wang$^{1,2,3}$,  and Hong-Yu Zhou$^{1,2,3}$}
\address{$^1$ Key Laboratory of Beam Technology and Material
Modification of Ministry of Education, Beijing Normal University,
Beijing 100875,  China\\
$^2$ College of Nuclear Science and Technology, Beijing Normal
University,
Beijing 100875, China\\
$^3$ Beijing Radiation Center, Beijing 100875,  China\\
$^4$ Department of Physics, Applied Optics Beijing Area Major
Laboratory, Beijing Normal University, Beijing 100875, China }
\date{\today }

\begin{abstract}
We present a scheme for multipartite entanglement purification of
quantum systems in a Greenberger-Horne-Zeilinger state with quantum
nondemolition detectors (QNDs). This scheme does not require the
controlled-not gates which cannot be implemented perfectly with
linear optical elements at present, but QNDs based on cross-Kerr
nonlinearities. It works with two steps, i.e., the bit-flipping
error correction and the phase-flipping error correction. These two
steps can be iterated perfectly with parity checks and simple
single-photon measurements. This scheme does not require the parties
to possess sophisticated single photon detectors. These features
maybe make this scheme more efficient and feasible than others in
practical applications.
\end{abstract}
\maketitle

\section{Introduction}

Entanglement plays an important role in quantum information
processing \cite{book}. For example, bipartite entangled states
provide some novel ways for quantum cryptography
\cite{Ekert91,BBM92,rmp,LongLiu,CORE}, quantum teleportation
\cite{teleportation} and quantum dense coding
\cite{densecoding,densecoding2,densecoding3}.  Multipartite
entangled states have many important applications in quantum
computation and quantum communication. It provides the superpower
of quantum computer \cite{book} and  the resource for quantum
error correction codes \cite{Calderbank}. Some important branches
of quantum  communication require multipartite entangled states to
set up the quantum channel, such as controlled teleportation
\cite{cteleportation,cteleportation2}, quantum secret sharing
\cite{QSS,QSS2,QSS3}  and quantum state sharing
\cite{QSTS,QSTS2,QSTS3,QSTS4,QSTS5}.

In experiment, the implement of quantum communication depends on
the transmission of quantum systems. However, the noise in quantum
channel will degrade the entanglement of the quantum system
transmitted, even make it in a mixed state, which will decrease
the fidelity of quantum teleportation  or make quantum
communication insecure. In this time,  the parties of quantum
communication usually exploit entanglement purification
\cite{Bennett1,Deutsch,Pan1,Simon,shengpra,Murao,Horodecki,Yong}
or  entanglement concentration
\cite{Bennett2,swapping1,swapping2,Yamamoto,zhao1,shengpra2} to
obtain some maximally entangled states from a less-entanglement
ensemble. Entanglement purification is used to increase the
entanglement of quantum systems in a mixed state, while
entanglement concentration is only used to obtain some maximally
entangled states from a set of pure entangled quantum systems .
The former is more general than the latter in a practical quantum
communication. By far, entanglement purification has been studied
not only for bipartite entangled quantum systems
\cite{Bennett1,Deutsch,Pan1,Simon,shengpra} but also for
multipartite entangled quantum systems
\cite{Murao,Horodecki,Yong}. As multipartite entanglement
purification is far more difficult than that for two-particle Bell
states, there are only serval multipartite entanglement
purification schemes \cite{Murao,Horodecki,Yong}, including that
for high-dimensional quantum systems.

In 1998,  Murao  et al.  \cite{Murao} presented a multipartite
entanglement purification protocol for  quantum systems in a
Greenberger-Horne-Zeilinger (GHZ) state with controlled-NOT (CNOT)
gates and local Hadamard transformations. Their protocol has been
generalized to high-dimensional multipartite quantum systems by
Cheong  et al. \cite{Yong} in 2007. In the latter \cite{Yong}, they
use some generalized XOR gates in high-dimensional systems, instead
of the common CNOT gates in two-dimensional systems, and the
Hadamard transformation is substituted by the quantum Fourier
transformation. It has been shown that with only single photon
sources and linear optical elements, the maximal probability for
achieving the CNOT gate is 3/4 \cite{Pittman}. So the CNOT gate
based on linear optics is beyond the reach of current technology.
These obstacles make the multipartite entanglement purification
protocols \cite{Murao,Horodecki,Yong} be hard to realize at present.

Cross-kerr nonlinearity provides a good tool to complete a
parity-check measurement \cite{QND1,QND2}. With quantum language,
the cross-Kerr nonlinearities can be described with the
Hamiltonian as follows \cite{QND1,QND2}:
\begin{eqnarray}
H_{ck}=\hbar\chi a^{\dagger}_{s}a_{s}a^{\dagger}_{p}a_{p},
\end{eqnarray}
where $a^{\dagger}_{s}$ and $a^{\dagger}_{p}$ denote the creation
operations, and  $a_{s}$ and $a_{p}$ are the annihilation
operations.  $\hbar\chi $ is the coupling strength of the
nonlinearity, which is decided by the property of material. For a
quantum signal in a Fock state with the form of
$|\Psi\rangle_s=c_{0}|0\rangle_{s}+c_{1}|1\rangle_{s}$
($|0\rangle_{s}$ and $|1\rangle_{s}$ denote that there are no
photon and one photon, respectively, in this state, and
$|c_{0}|^{2}+|c_{1}|^{2}=1$) and a coherent probe beam in the
state $|\alpha\rangle_{p}$, after the interaction with the
cross-Kerr nonlinear medium the whole system evolves as
\begin{eqnarray}
U_{ck}|\Psi\rangle_{s}|\alpha\rangle_{p}&=&
e^{iH_{ck}t/\hbar}[c_{0}|0\rangle_{s}+c_{1}
|1\rangle_{s}]|\alpha\rangle_{p} \nonumber\\
&=& c_{0}|0\rangle_{s}|\alpha\rangle_{p}+c_{1}|1\rangle_{s}|\alpha
e^{i\theta}\rangle_{p},
\end{eqnarray}
where $\theta=\chi t$ and $t$ is the interaction time. It is shown
that the coherent beam picks up a phase shift proportional to the
number of the photons in the Fock state.

In this paper, we present a feasible scheme for multipartite
entanglement purification of  quantum systems in a GHZ state by
constructing nondestructive quantum nondemolition detectors (QND)
with cross-Kerr nonlinearities. The task of multipartite
entanglement purification can be completed with two steps which can
be iterated perfectly. The first one is to purify the bit-flipping
errors in multipartite quantum systems, and the second one is to
purify their phase-flipping errors. This protocol does not require
the CNOT gate based on linear optics and sophisticated single-photon
detectors, which makes it more feasible in practical applications.
Moreover, it has the same yield as that with CNOT gates but reduces
a large number of quantum resources in principle.

\section{multipartite entanglement purification with quantum nondemolition detectors}

\subsection{quantum nondemolition detector and description of errors}

The principle of our nondestructive quantum nondemolition detector
(QND) is shown in Fig.1. It is composed of two cress-Kerr
nonlinearities ($ck_1$ and $ck_2$), four polarization beam splitters
(PBSs), a coherent beam $\vert \alpha\rangle_p$, and an X homodyne
measurement. $b_1$ and $b_2$ represent the up spatial mode and the
down spatial mode, respectively.  Each polarization beam splitter
(PBS) is used to pass through the horizontal polarization photons
$|H\rangle$ and reflect the vertical polarization photons
$|V\rangle$. The cross-Kerr nonlinearity will make the coherent beam
$\vert \alpha \rangle_p$ pick up a phase shift $\theta$ if there is
a photon in the mode. The probe beam $\vert \alpha \rangle_p$ will
pick up a phase shift $\theta$ if the state of the two photons
injected into the two spatial modes $b_1$ and $b_2$ is $|HH\rangle$
or $|VV\rangle$; otherwise it picks up a phase shift $0$ (for
$|VH\rangle$) or $2\theta$ (for $|HV\rangle$). That is, when the
parity of the two photons is even, the coherent beam $\vert \alpha
\rangle_p$ will pick up a phase shift $\theta$; otherwise it will
pick up $0$ or $2\theta$. Each party of quantum communication can
determine the parity of his two photons with an X homodyne
measurement. With this QND, we can distinguish superpositions and
mixtures of $|HH\rangle$ and $|VV\rangle$ from $|HV\rangle$ and
$|VH\rangle$.

\begin{figure}[!h]
\begin{center}
\includegraphics[width=7cm,angle=0]{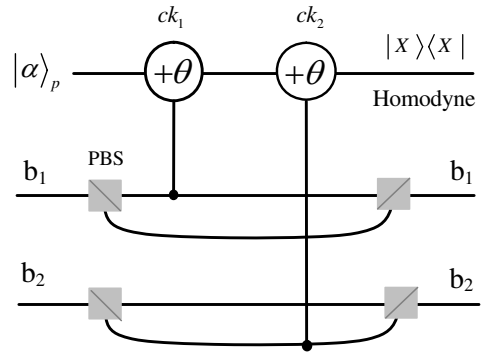}
\caption{Schematic diagram showing the principle of our
nondestructive quantum nondemolition detector (QND). }
\end{center}
\end{figure}

A multipartite GHZ state for spin 1/2 systems can be written as
\begin{eqnarray}
|\phi^{+}\rangle_s=\frac{1}{\sqrt{2}}(|00\cdots0\rangle +
|11\cdots1\rangle).
\end{eqnarray}
Here $\vert 0\rangle \equiv \vert H\rangle$ and $\vert 1\rangle
\equiv \vert V\rangle$ represent the horizontal polarization state
and the vertical polarization one, respectively. They are the two
eigenvectors of the basis $Z$. In the following, we first use
three-particle GHZ-state systems as an example for demonstrating the
principle of our multipartite entanglement purification scheme and
then discuss the case for $N$-particle systems in a GHZ state. This
scheme includes two steps: one for the bit-flipping error correction
and the other for phase-flipping error correction. As this scheme
works with quantum nondemolition detectors, instead of CNOT gates,
we denote it QND scheme.

There are eight three-particle GHZ states, i.e.,
\begin{eqnarray}
|\Phi^{\pm}\rangle_{ABC}=\frac{1}{\sqrt{2}}(|000\rangle\pm|111\rangle)_{ABC},\nonumber\\
|\Phi_{1}^{\pm}\rangle_{ABC}=\frac{1}{\sqrt{2}}(|100\rangle\pm|011\rangle)_{ABC},\nonumber\\
|\Phi_{2}^{\pm}\rangle_{ABC}=\frac{1}{\sqrt{2}}(|010\rangle\pm|101\rangle)_{ABC},\nonumber\\
|\Phi_{3}^{\pm}\rangle_{ABC}=\frac{1}{\sqrt{2}}(|001\rangle\pm|110\rangle)_{ABC}.\label{GHZstate}
\end{eqnarray}
Here the subscripts $A$, $B$, and $C$ represent the particles
belonging to the three parties, say Alice, Bob, and Charlie,
respectively. Suppose that the original GHZ state transmitted
among the three parties is $|\Phi^{+}\rangle_{ABC}$. If a
bit-flipping error takes place on the particle in this GHZ state
after it is transmitted in a noisy channel, the three-particle
system is in the state $|\Phi^{+}_{1}\rangle_{ABC}$. We label that
a bit-flipping error occurs on the first particle. If
$|\Phi^{+}\rangle$ becomes $|\Phi^{-}\rangle$, there is a
phase-flipping error. Sometimes, both a bit-flipping error and a
phase-flipping error will take place on a three-particle quantum
system transmitted in a noisy channel. The task for purifying
three-particle entangled systems requires to correct both
bit-flipping errors and phase-flipping errors on the quantum
systems.

\subsection{bit-flipping error correction}
\label{bfec}

Suppose that Alice, Bob and Charlie share an ensemble $\rho$ after
the transmission of particles, i.e.,
\begin{eqnarray}
\rho=F|\Phi^{+}\rangle\langle\Phi^{+}|+(1-F)|\Phi_{1}^{+}\rangle\langle\Phi_{1}^{+}|.\label{ensemblerho}
\end{eqnarray}
It means that there is a bit-flipping error on the quantum system
with a probability of $1-F$. Here $F (>\frac{1}{2})$ is the fidelity
of the quantum systems transmitted. For correcting this error, the
three parties divide their quantum systems in the ensemble $\rho$
into many groups and each groups is composed of a pair of
three-photon quantum systems, same as the first multipartite
entanglement purification protocol by Murao  et al. \cite{Murao}  in
1998. We label each group with $A_1B_1C_1A_2B_2C_2$ (the two
three-photon quantum systems $A_1B_1C_1$ and $A_2B_2C_2$).

\begin{figure}[!h]
\begin{center}
\includegraphics[width=6cm,angle=0]{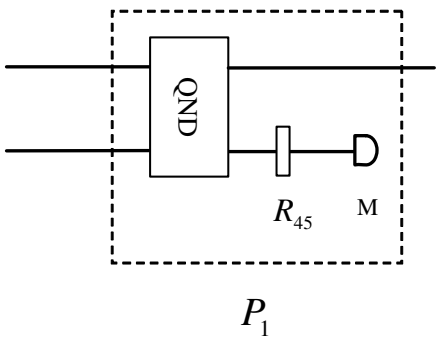}
\caption{The principle of the bit-flipping error correction with
QND. The $45^{\circ}$ wave plate $R_{45}$ is used to transform the
state $|0\rangle$ to $1/\sqrt{2}(|0\rangle+|1\rangle)$ and
$|1\rangle$ to $1/\sqrt{2}(|0\rangle-|1\rangle)$. $M$ represents a
single-photon measurement with the basis $Z$. We denote this
process $P_{1}.$ }
\end{center}
\end{figure}

The principle of our scheme for correcting a bit-flipping error is
shown in Fig.2. The state of the two quantum systems
$A_1B_1C_1A_2B_2C_2$ can be viewed as the mixture of four pure
states, i.e., $|\Phi^{+}\rangle\otimes|\Phi^{+}\rangle$ with a
probability of $F^{2}$, both
$|\Phi^{+}\rangle\otimes|\Phi^{+}_{1}\rangle$ and
$|\Phi^{+}_{1}\rangle\otimes|\Phi^{+}\rangle$ with an equal
probability of $F(1-F)$, and
$|\Phi^{+}_{1}\rangle\otimes|\Phi^{+}_{1}\rangle$ with a
probability of $(1-F)^{2}$. For each group, Alice takes her two
photons $A_1A_2$ to pass through the setup shown in Fig.2. The
photon $A_1$ entrances the up spatial mode and the photon $A_2$
entrances the down spatial mode. So do the other parties Bob and
Charlie. After the QNDs, the three parties compare the parity of
their photons. They only keep the groups for which all the three
parties get an even parity. After these operations, the quantum
systems are in a new mixed state which is composed of the two
states
\begin{eqnarray}
|\phi\rangle=\frac{1}{\sqrt{2}}(|000000\rangle+|111111\rangle)_{A_1B_1C_1A_2B_2C_2}
\end{eqnarray}
with a probability of $\frac{1}{2}F^{2}$ and
\begin{eqnarray}
|\phi_{1}\rangle=\frac{1}{\sqrt{2}}(|100100\rangle+|011011\rangle)_{A_1B_1C_1A_2B_2C_2}
\end{eqnarray}
with a probability of $\frac{1}{2}(1-F)^2$ as the two
cross-combinations $|\Phi^{+}\rangle\otimes|\Phi^{+}_{1}\rangle$
and $|\Phi^{+}_{1}\rangle\otimes|\Phi^{+}\rangle$ never lead  all
the three parties to have the same parity.

After the rotation $R_{45}$ on each photon in the down spatial mode,
the three parties measure their photons out of the down spatial
modes with the basis $Z$. The wave plate $R_{45}$ is used to rotate
the horizontal and vertical polarizations by $45^{\circ}$ (this task
can be completed by a half wave plate whose orientation is
22.5$^\circ$), i.e., it acts as a Hadamard (H) gate,
\begin{eqnarray}
|0\rangle & \rightarrow & \frac{1}{\sqrt{2}}(|0\rangle+|1\rangle),\\
|1\rangle & \rightarrow & \frac{1}{\sqrt{2}}(|0\rangle-|1\rangle).
\end{eqnarray}
After the rotations, $|\phi\rangle$ becomes
\begin{eqnarray}
|\phi\rangle^{'}=\frac{1}{\sqrt{2}}(|000\rangle(\frac{1}{\sqrt{2}})^{\otimes3}(|0\rangle+|1\rangle)^{\otimes3}\nonumber\\
+|111\rangle(\frac{1}{\sqrt{2}})^{\otimes3}(|0\rangle-|1\rangle)^{\otimes3})
\end{eqnarray}
and $|\phi_{1}\rangle$ becomes
\begin{eqnarray}
|\phi_1\rangle^{'}=\frac{1}{\sqrt{2}}(|100\rangle(\frac{1}{\sqrt{2}})^{\otimes3}(|0\rangle-|1\rangle)(|0\rangle
+ |1\rangle)^{\otimes2}\nonumber\\
+|011\rangle(\frac{1}{\sqrt{2}})^{\otimes3}(|0\rangle+|1\rangle)(|0\rangle-|1\rangle)^{\otimes2}).
\end{eqnarray}
After the measurements on the photons  $A_2$, $B_2$, and $C_2$, the
three parties will obtain the state $|\Phi^{+}\rangle_{A_1B_1C_1}$
with a fidelity of $F'=\frac{F^{2}}{F^{2}+(1-F)^{2}}>F$ if their
outcome is $|000\rangle_{A_2B_2C_2}$, $|011\rangle_{A_2B_2C_2}$,
$|101\rangle_{A_2B_2C_2}$ or $|110\rangle_{A_2B_2C_2}$. If their
outcome is $|001\rangle$, $|010\rangle$, $|100\rangle$ or
$|111\rangle$, they need only to flip the phase of the quantum
systems kept and get the same result above. This task can be
accomplished with a $90^{\circ}$ rotation on a photon in each
three-photon quantum system. In detail, one of the three parties let
his photon pass through a half-wave plate whose orientation is
90$^\circ$.

In the process above, each party chooses the phase shift $\theta$
for purification and gets the even parity $|HH\rangle$ or
$|VV\rangle$. Another outcome for each one of
$|\Phi^{+}\rangle\otimes|\Phi^{+}\rangle$ is an odd parity, i.e.,
$|000111\rangle_{A_1B_1C_1A_2B_2C_2}$ or
$|111000\rangle_{A_1B_1C_1A_2B_2C_2}$. The two photons of each
party are in the state $|HV\rangle$ or $|VH\rangle$, which leads
to a phase shift $2\theta$ or $0$, respectively. These photons are
discarded in the  process above. In this way, the yield of this
scheme is half of the original protocol proposed by Murao \emph{et
al.} \cite{Murao} with CNOT gates. However, if we choose a proper
material for the QND and make $\theta=\pi$, we cannot distinguish
the phase shift $2\theta$ and $0$. In this time, the three parties
can also keep their photons when they all get an odd parity. The
state of the six photons becomes
\begin{eqnarray}
|\phi\rangle^o=\frac{1}{\sqrt{2}}(|000111\rangle+|111000\rangle)_{A_1B_1C_1A_2B_2C_2}
\end{eqnarray}
with a probability of $\frac{1}{2}F^{2}$ and
\begin{eqnarray}
|\phi_{1}\rangle^o=\frac{1}{\sqrt{2}}(|001110\rangle+|110001\rangle)_{A_1B_1C_1A_2B_2C_2}
\end{eqnarray}
with a probability of $\frac{1}{2}(1-F)^{2}$. With a bit-flipping
operation on each photon in the quantum system $A_2B_2C_2$, we get
the same result as that with an even parity, which will double the
yield.

For correcting the bit-flipping errors in multipartite entangled
quantum systems, we can follow the same step of that for the
three-particle  GHZ-state systems.  We should only  increase the
number of  the QND equipments. For instance, for  $N$-particle
quantum systems whose original states are
\begin{eqnarray}
|\Phi^+\rangle_{N}=\frac{1}{\sqrt{2}}(|00\cdots 0\rangle+|11\cdots
1\rangle)\label{stateN},
\end{eqnarray}
if a bit-flipping error occurs on the first particle,  the $N$
parties chooses the same phase shift $\theta$ after their QNDs and
the state of a group of the quantum systems kept (two GHZ-state
quantum systems) becomes
\begin{eqnarray}
|\phi\rangle_{2N}=\frac{1}{\sqrt{2}}(|000\cdots000\rangle+|111\cdots111\rangle)\label{correctN}
\end{eqnarray}
with a probability of $\frac{1}{2}F^{2}$ and
\begin{eqnarray}
|\phi_{1}\rangle_{2N}=\frac{1}{\sqrt{2}}(|10\cdots010\cdots0\rangle+|01\cdots101\cdots1\rangle)\label{errorN}
\end{eqnarray}
with a probability of $\frac{1}{2}(1-F)^{2}$. After a $45^{\circ}$
rotation on each photon in the second $N$-particle quantum system,
Eq.(\ref{correctN}) becomes
\begin{eqnarray}
|\phi\rangle^{'}_{2N}=\frac{1}{\sqrt{2}}(|00\cdots0\rangle(\frac{1}{\sqrt{2}})^{\otimes N}(|0\rangle
 + |1\rangle)^{\otimes N}\nonumber\\
+|11\cdots1\rangle(\frac{1}{\sqrt{2}})^{\otimes
N}(|0\rangle-|1\rangle)^{\otimes N})
\end{eqnarray}
and Eq.(\ref{errorN}) becomes
\begin{eqnarray}
|\phi\rangle^{'}_{2N}=\frac{1}{\sqrt{2}}(|10\cdots0\rangle(\frac{1}{\sqrt{2}})^{\otimes N}(|0\rangle
 - |1\rangle)(|0\rangle+|1\rangle)^{\otimes(N-1)}\nonumber\\
+|01\cdots1\rangle(\frac{1}{\sqrt{2}})^{\otimes
N}(|0\rangle+|1\rangle)(|0\rangle-|1\rangle)^{\otimes(N-1)}).\nonumber\\
\end{eqnarray}
After the measurements on the photons in the second quantum system
with the basis $Z$, we will get the state $|\Phi^{+}\rangle_{N}$
with a fidelity of $\frac{F^{2}}{F^{2}+(1-F)^{2}}$  if the number
of $|1\rangle$ is even; otherwise,, we get the
$|\Phi^{-}\rangle_{N}$ with the same fidelity if the number of the
outcomes $|1\rangle$ is odd. Here
\begin{eqnarray}
|\Phi^-\rangle_{N}=\frac{1}{\sqrt{2}}(|00\cdots 0\rangle -
|11\cdots 1\rangle)\label{stateN2}.
\end{eqnarray}
With a phase-flipping operation, the state $|\Phi^{-}\rangle_{N}$
is transformed into the state $|\Phi^{+}\rangle_{N}$.

In essence, the process above is used to purify the bit-flipping
error occurring on the first particle. Those on the other particles
can also be corrected in the same way and one will get the same
result above easily. That is, they can correct the bit-flipping
errors in the state $\rho_N$, here
\begin{eqnarray}
\rho_N &=& F_0|\Phi^{+}\rangle\langle\Phi^{+}|
+F_1|\Phi_{1}^{+}\rangle\langle\Phi_{1}^{+}|
+F_2|\Phi_{2}^{+}\rangle\langle\Phi_{2}^{+}|\nonumber\\
&+& \cdots
+F_N|\Phi_{N}^{+}\rangle\langle\Phi_{N}^{+}|.\label{ensemblerhoN}
\end{eqnarray}

\subsection{phase-flipping error correction}

A phase-flipping error cannot be corrected directly, different from
a bit-flipping error, but it can be transformed into a bit-flipping
error with H operations. With a H operation on each photon, the
states shown in Eq. (\ref{GHZstate}) are transformed into the
following ones
\begin{eqnarray}
|\Psi^{+}\rangle=\frac{1}{2}(|000\rangle+|011\rangle+|101\rangle+|110\rangle),\nonumber\\
|\Psi^{-}\rangle=\frac{1}{2}(|001\rangle+|010\rangle+|100\rangle+|111\rangle),\nonumber\\
|\Psi_{1}^{+}\rangle=\frac{1}{2}(|000\rangle+|011\rangle-|101\rangle-|110\rangle),\nonumber\\
|\Psi_{1}^{-}\rangle=\frac{1}{2}(|001\rangle+|010\rangle-|100\rangle-|111\rangle),\nonumber\\
|\Psi_{2}^{+}\rangle=\frac{1}{2}(|000\rangle-|011\rangle+|101\rangle-|110\rangle),\nonumber\\
|\Psi_{2}^{-}\rangle=\frac{1}{2}(|001\rangle-|010\rangle+|100\rangle-|111\rangle),\nonumber\\
|\Psi_{3}^{+}\rangle=\frac{1}{2}(|000\rangle-|011\rangle-|101\rangle+|110\rangle),\nonumber\\
|\Psi_{3}^{-}\rangle=\frac{1}{2}(|001\rangle-|010\rangle-|100\rangle+|111\rangle).\label{phaseflipstate}
\end{eqnarray}

\begin{figure}[!h]
\begin{center}
\includegraphics[width=7cm,angle=0]{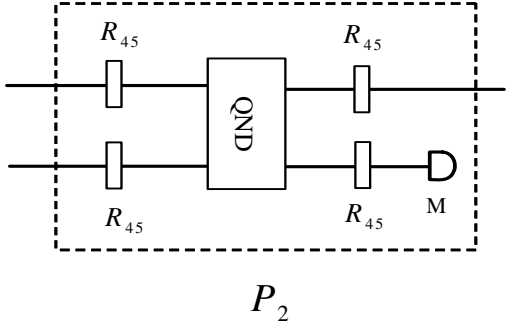}
\caption{The principle of the phase-flipping error correction with
QND. We denote this process $P_{2}.$}
\end{center}
\end{figure}

From Eq.(\ref{phaseflipstate}), one can see that the
transformation between phase-flipping errors and bit-flipping
errors in three-particle GHZ states is more complex than that in
Bell states \cite{Bennett1,Deutsch,Pan1,Simon,shengpra}. We can
not use the equipment shown in Fig.2 to purify the states in
Eq.(\ref{phaseflipstate}) directly. That is, we cannot exploit
simply H operations to complete the transformation between
phase-flipping errors and bit-flipping errors in three-particle
GHZ states perfectly, different from Bell states. Fortunately, the
eight states can be divided into two groups. In the GHZ states
with the superscript $+$, the number of $|1\rangle$ is even such
as $|\Psi^{+}\rangle$, $|\Psi_{1}^{+}\rangle$,
$|\Psi_{2}^{+}\rangle$, and $|\Psi_{3}^{+}\rangle$. In the other
group, the number of $|1\rangle$ is odd such as
$|\Psi^{-}\rangle$, $|\Psi_{1}^{-}\rangle$,
$|\Psi_{2}^{-}\rangle$, and $|\Psi_{3}^{-}\rangle$. The result is
also valid for multipartite entangled states ($N>3$). With
grouping, we can complete the task of phase-flipping error
correction.

Now, we use a pair of partner states $|\Psi^{+}\rangle$ and
$|\Psi^{-}\rangle$ as an example to describe the principle of the
phase-flipping error correction in our scheme. The density matrix
of the ensemble can be written as
\begin{eqnarray}
\rho'=F|\Psi^{+}\rangle\langle\Psi^{+}|+(1-F)|\Psi^{-}\rangle\langle\Psi^{-}|.\label{ensemblerho2}
\end{eqnarray}
For each pair of the entangled quantum systems picked out from
this ensemble $A_1B_1C_1$ and $A_2B_2C_2$, their state can be
viewed as the mixture of four pure states, i.e., $|\Psi^{+}\rangle
\otimes |\Psi^{+}\rangle$, $|\Psi^{+}\rangle \otimes
|\Psi^{-}\rangle$, $|\Psi^{-}\rangle \otimes |\Psi^{+}\rangle$,
and $|\Psi^{-}\rangle \otimes |\Psi^{-}\rangle$, similar to the
case for the bit-flipping error correction. After passing through
the QND of $P_{2}$ shown in Fig.3, the three parties measure the
phase shifts of their coherent beams with X homodyne measurements
and keep the six photons if  their phase shifts all are $\theta$;
otherwise, they discard their six photons. By choosing the samples
with even parities,  the cross-combinations
$|\Psi^{+}\rangle\otimes |\Psi^{-}\rangle$ and $|\Psi^{-}\rangle
\otimes |\Psi^{+}\rangle$  will never appear. The remaining items
are
\begin{eqnarray}
|\varphi\rangle=\frac{1}{2}(|000000\rangle+|011011\rangle+|101101\rangle+|110110\rangle) \nonumber\\
\label{evenparity1}
\end{eqnarray}
with a probability of $\frac{1}{2}F^{2}$  and
\begin{eqnarray}
|\varphi'\rangle=\frac{1}{2}(|001001\rangle+|010010\rangle+|100100\rangle+|111111\rangle) \nonumber\\
\label{evenparity2}
\end{eqnarray}
with a probability of $\frac{1}{2}(1-F)^{2}$.
Eq.(\ref{evenparity1}) and Eq.(\ref{evenparity2}) are both
six-photon entangled states. In order to get the three-photon GHZ
state $|\Psi^{+}\rangle$, each party needs to measure this photon
out of the low spatial mode of the setup $P_{2}$ with the basis
$X=\{\frac{1}{\sqrt{2}}(\vert 0\rangle + \vert 1\rangle),
\frac{1}{\sqrt{2}}(\vert 0\rangle - \vert 1\rangle)\}$. That is,
the three parties rotate the three photons $A_2$, $B_2$ and $C_2$
by $45^{\circ}$, which will complete the transformations
\begin{widetext}
\begin{center}
\begin{eqnarray}
|\varphi\rangle &\rightarrow& \frac{1}{4\sqrt{2}}
|000\rangle_{A_2B_2C_2}(|000\rangle+|001\rangle+|010\rangle+|011\rangle
+|100\rangle+|101\rangle+|110\rangle+|111\rangle)_{A_1B_1C_1}\nonumber\\
&& \;\;\;\;\; + |011\rangle_{A_2B_2C_2}(|000\rangle-|001\rangle-|010\rangle+|011\rangle
+|100\rangle-|101\rangle-|110\rangle+|111\rangle)_{A_1B_1C_1}\nonumber\\
&& \;\;\;\;\; + |101\rangle_{A_2B_2C_2}(|000\rangle-|001\rangle+|010\rangle-|011\rangle
-|100\rangle+|101\rangle-|110\rangle+|111\rangle)_{A_1B_1C_1}\nonumber\\
&& \;\;\;\;\; + |110\rangle_{A_2B_2C_2}(|000\rangle+|001\rangle-|010\rangle-|011\rangle
-|100\rangle-|101\rangle+|110\rangle+|111\rangle)_{A_1B_1C_1} \label{finalstate1} \nonumber \\
\end{eqnarray}
\end{center}
\end{widetext}
\begin{widetext}
\begin{center}
\begin{eqnarray}
|\varphi\rangle' & \rightarrow &\frac{1}{4\sqrt{2}}
|001\rangle_{A_2B_2C_2}(|000\rangle-|001\rangle+|010\rangle-|011\rangle
+|100\rangle-|101\rangle+|110\rangle-|111\rangle)_{A_1B_1C_1}\nonumber\\
&& \;\;\;\;\; + |010\rangle_{A_2B_2C_2}(|000\rangle+|001\rangle-|010\rangle-|011\rangle
+|100\rangle+|101\rangle-|110\rangle-|111\rangle)_{A_1B_1C_1}\nonumber\\
&& \;\;\;\;\; + |100\rangle_{A_2B_2C_2}(|000\rangle+|001\rangle+|010\rangle+|011\rangle
-|100\rangle-|101\rangle-|110\rangle-|111\rangle)_{A_1B_1C_1}\nonumber\\
&& \;\;\;\;\; + |111\rangle_{A_2B_2C_2}(|000\rangle-|001\rangle-|010\rangle+|011\rangle
-|100\rangle+|101\rangle+|110\rangle-|111\rangle)_{A_1B_1C_1}, \label{finalstate2}  \nonumber\\
\end{eqnarray}
\end{center}
\end{widetext}
and then they measure their photons $A_2$, $B_2$ and $C_2$ with
the basis $Z$.

From Eqs. (\ref{finalstate1}) and (\ref{finalstate2}), one can see
that the three parties will get the state
$|\Psi^{+}\rangle_{A_1B_1C_1}$ with some unitary operations if the
outcome of the measurements on the three particles $A_2$, $B_2$,
and $C_2$ is $|000\rangle_{A_2B_2C_2}$, $|011\rangle_{A_2B_2C_2}$,
$|101\rangle_{A_2B_2C_2}$ or $|110\rangle_{A_2B_2C_2}$, which
takes place with a probability of $\frac{1}{2}F^{2}$; otherwise,
they get the state $|\Psi^{-}\rangle_{A_1B_1C_1}$ with a
probability of $\frac{1}{2}(1-F)^{2}$. The three parties can
transform the states $|\Psi^{+}\rangle$ and $|\Psi^{-}\rangle$
into $|\Phi^{+}\rangle$ and $|\Phi^{-}\rangle$, respectively, by
adding a Hadamard transformation ($45^{\circ}$ rotations) on each
photon in the first three-photon quantum system $A_1B_1C_1$. In
this way, they will get a new mixed entangled state with the
fidelity of $\frac{F^{2}}{F^{2}+(1-F)^{2}}$, same as that for
bit-flipping error correction.

In essence, in the process of purifying the phase-flipping error in
the quantum systems, the parties of quantum communication first
transform the phase-flipping errors into the bit-flipping errors and
then correct them by comparing their parities. Although the
transformation on multi-particle GHZ states make them more complex
than Bell states, the principles of the bit-flipping error
correction and the phase-flipping error correction for
multi-particle GHZ states with parity checks are similar to those
for Bell states \cite{shengpra}. The difference is just that Bell
states are more symmetrical than multi-particle GHZ states under a
Hadamard transformation on each particle, which makes the
entanglement purification of Bell states easier than that of
multi-particle GHZ states. Same as the bit-flipping error
correction, the parties can also exploit proper QNDs to improve
their yield.

\section{multipartite entanglement purification  with polarizing beam splitters}

So far, there are mainly three types of principles for the
entanglement purification of two-photon Bell states. One is based
on CNOT gates, which is the pioneer for entanglement purification.
The second one is based on PBSs, which is more feasible than the
first one at present if it is used to improve partially the
entanglement of the entangled quantum systems transmitted although
its yield is in principle half of that with CNOT gates. The third
type is based on parity checks with cross-Kerr nonlinearities. It
is more feasible than the first type and has the same yield as the
latter. We will show that these differences exist for the
entanglement purification of multi-particle GHZ states yet.

\begin{figure}[!h]
\begin{center}
\includegraphics[width=6cm,angle=0]{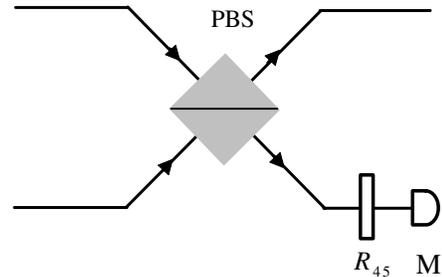}
\caption{The principle of the bit-flipping error correction on
three-particle GHZ states with PBSs and sophisticated
single-photon detectors, similar to the Pan's protocol
\cite{Pan1}. Each party has a setup for the bit-flipping error
correction, and the three parties choose the six-mode events to
ensure that they all get an even parity by classical
communication}
\end{center}
\end{figure}

The principle of the entanglement purification of  multipartite
entangled quantum systems with PBSs is similar to that with QNDs
discussed above. We call it MPBS protocol. It also contains two
steps: a bit-flipping error correction ($P_1$) and a
phase-flipping error correction ($P_2$).  We describe the first
step with an example of purifying an ensemble $\rho$ shown in Eq.
(\ref{ensemblerho}) and the second step with $\rho'$ shown in Eq.
(\ref{ensemblerho2}) below. The multipartite entanglement
purification for other cases are the same as that for an ensemble
$\rho$ with or without a little modification.

\begin{figure}[!h]
\begin{center}
\includegraphics[width=6cm,angle=0]{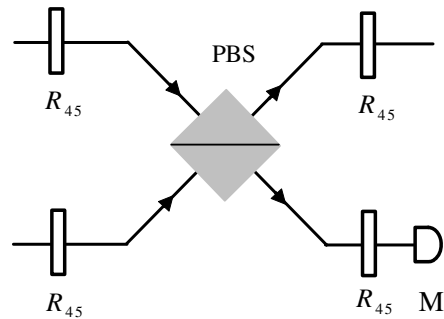}
\caption{The principle of the phase-flipping error correction on
three-particle GHZ states with PBSs.}
\end{center}
\end{figure}

The principle of the bit-flipping error correction on
three-particle GHZ states with PBSs is shown in Fig.4. Each party
of quantum communication first lets his two photons coming from
two quantum systems ($A_1B_1C_1$ and $A_2B_2C_2$) pass through the
setup shown in Fig.4 from the two spatial modes, respectively, and
then they
 pick up the events in which there is one and only
one photon in each spatial mode (call it a six-mode event),
similar to the four-mode events in Ref.\cite{Pan1}. The quantum
systems kept are in the mixture of the state
\begin{eqnarray}
|\phi\rangle=\frac{1}{\sqrt{2}}(|000000\rangle+|111111\rangle)_{A_1B_1C_1A_2B_2C_2}
\end{eqnarray}
with a probability of $\frac{1}{2}F^{2}$ and
\begin{eqnarray}
|\phi_{1}\rangle=\frac{1}{\sqrt{2}}(|100100\rangle+|011011\rangle)_{A_1B_1C_1A_2B_2C_2}
\end{eqnarray}
with a probability of $\frac{1}{2}(1-F)^2$, as same as the case in
which all the three parties get an even parity in the entanglement
purification of the bit-flipping errors with QNDs in Sec.
\ref{bfec}. In this way, the three parties can get a new ensemble
of the fidelity of $F'$ with some single-photon measurements and
unitary operations, same as that with QNDs. The phase-flipping
error correction can also be accomplished with the setup shown in
Fig.5 by picking up only the six-mode events. The other processes
are same as those with QNDs except for exploiting sophisticated
single-photon detectors to distinguish the six-mode events from
others.

\section{Discussion and summary}

In our scheme, we detail the multipartite entanglement
purification with two steps: one is the bit-flipping error
correction and the other is the phase-flipping error correction.
When entangled qubits are transmitted in a practical channel, it
is possible to take place both a bit-flipping error and a
phase-flipping error on qubits. A simple example is called the
"Werner-type" state \cite{Murao}, i.e,
\begin{eqnarray}
\rho_{w}=x|\phi^{+}\rangle\langle\phi^{+}|+\frac{1-x}{2^{N}}\emph{1}.
\end{eqnarray}
Its fidelity  is
\begin{eqnarray}
f_{\rho}=\langle\phi^{+}|\rho_{w}|\phi^{+}\rangle=x+(1-x)/2^{N}.
\end{eqnarray}
If we want to purify this "Werner-type" state for obtaining the
maximally entangled state $|\phi^{+}\rangle$,  the bit-flipping
error correction  $P_{1}$ and the phase-flipping error correction
$P_{2}$ are both needed. In the Murao's protocol \cite{Murao},
they use $P_{1}+P_{2}$ to purify this mixed state. That is, the
whole process for purification should be
$P_{1}P_{2}P_{1}P_{2}P_{1}\ldots$. They found that the
$P_{1}+P_{2}$ protocol is not optimal for two-particle quantum
systems, so it may be not optimal for the purification of
multipartite entangled systems. In our protocol, we do not use the
$P_{1}+P_{2}$ process, the order of $P_{1}$ and $P_{2}$ is
arbitrary. However, we do not know which order is the optimal one.
The purification of a "Werner-type" state is more complicated than
that with a single error correction, a bit-flipping error or a
phase-flipping one. We cannot get a deterministic expression for
describing the iteration of the fidelity of ensembles kept like
those in Refs. \cite{Bennett1,Deutsch}. It should be studied with
some numerical methods according to the noise of the channel.

In the MPBS protocol, the parties only keep the events in which
each spatial mode has one and only one photon, which requires each
party to possess at least a sophisticated single-photon detector.
At present, sophisticated single-photon detectors are not
feasible. However, cross-Kerr nonlinearities provide a good way
for the parity check of the polarization states of two photons.
This feature can be used to construct a QND for entanglement
purification of multipartite quantum systems. In our QND protocol,
the QND acts as not only the role of a CNOT gate but also that of
a photon-number detector, which makes the process for entanglement
purification can be iterated perfectly. As it dose not require a
CNOT gate with linear optical elements and sophisticated
single-photon detectors,  this protocol is more convenient than
the MPBS protocol in practical applications.

Certainly, cross-Kerr nonlinear can be used to construct a CNOT gate
\cite{QND1}. With CNOT gates, multipartite entanglement purification
can be completed with the Murao's protocol \cite{Murao}. In fact,
parity check is enough for entanglement purification \cite{Pan1} and
it requires less quantum resources largely, compared with CNOT gates
\cite{QND1}. In our scheme, we use cross-Kerr nonlinear to construct
a QND which acts as the role of parity check and photon number
detector. One can also exploit other QNDs \cite{QND3,QND4,QND5} to
accomplish these tasks.

Same as all existing multipartite entanglement purification
protocols \cite{Murao,Horodecki,Yong} (as well as the entanglement
purification protocols for two-particle systems
\cite{Bennett1,Deutsch,Pan1,Simon,shengpra}), our scheme requires
the parties possess the capability of storing the entangled photons
in principle for improving the fidelity of multipartite GHZ-state
quantum systems. At present, storing a quantum state for a long time
is not easy. The parties can exploit optical delays to complete the
task of storing a quantum state for a short time. On the other hand,
our scheme requires some Hadamard gates on photons, same as the
first multipartite entanglement purification protocol by  Murao et
al. \cite{Murao}. Of course, the task is in principle not difficult
to be accomplished with a half-wave plate whose orientation is
22.5$^\circ$.

In summary, we have presented a multipartite entanglement
purification protocol for quantum systems in a GHZ state. The task
of entanglement purification can be accomplished with two steps. The
first step is to decrease the rate of bit-flipping errors and the
other is used for phase-flipping errors. In our protocol, we use a
QND to check the parity of the polarization states of two photons.
Each QND detector acts as the role of both a CNOT gate and a
photon-number detector, which makes this protocol feasible for the
iteration of purification. With a weak cross-Kerr medium, the
parties of quantum communication can keep the events in which they
all get an even parity for a pair of multipartite entangled quantum
systems. In this time, this protocol has the same yield as that with
PBSs and sophisticated single-photon detectors. If the parties can
choose a proper cross-Kerr nonlinear medium and a strong coherent
beam, they can also exploit the events in which they all get an odd
parity. With this modification, this protocol has the same yield as
that with CNOT gates. Compared with the Murao's protocol
\cite{Murao}, this protocol provides a practical way to realize
entanglement purification of multipartite entangled quantum systems
which are very useful in a long-distance quantum communication.

\section*{ACKNOWLEDGEMENTS}

This work is supported by the National Natural Science Foundation
of China under Grant No. 10604008, A Foundation for the Author of
National Excellent Doctoral Dissertation of P. R. China under
Grant No. 200723 and  Beijing Natural Science Foundation under
Grant No. 1082008.

\end{document}